\documentclass[sigconf]{acmart}

\usepackage{booktabs} 
\usepackage{todonotes}
\usepackage{subcaption}
\usepackage{xcolor}
\usepackage{graphbox}
\usepackage{natbib}
\usepackage{listings}
\usepackage{hyperref}
\usepackage{dirtree}

\setcopyright{none}
\renewcommand\footnotetextcopyrightpermission[1]{}
\begin{document}

\fancyhead{}
\title{Insights on the V3C2 Dataset}

\author{Luca Rossetto}
\orcid{0000-0002-5389-9465}
\affiliation{%
  \institution{University of Zurich}
  \streetaddress{Zurich, Switzerland}
}
\email{rossetto@ifi.uzh.ch}

\author{Abraham Bernstein}
\orcid{0000-0002-0128-4602}
\affiliation{%
  \institution{University of Zurich}
  \streetaddress{Zurich, Switzerland}
}
\email{bernstein@ifi.uzh.ch}

\author{Klaus Schoeffmann}
\orcid{0000-0002-9218-1704}
\affiliation{%
    \institution{University of Klagenfurt}
    \streetaddress{Klagenfurt, Austria}
}
\email{ks@itec.aau.at}

\begin{abstract}
For research results to be comparable, it is important to have common datasets for experimentation and evaluation. The size of such datasets, however, can be an obstacle to their use. The Vimeo Creative Commons Collection (V3C) is a video dataset designed to be representative of video content found on the web, containing roughly $3\,800$ hours of video in total, split into three shards. 

In this paper, we present insights on the second of these shards (V3C2) and discuss their implications for research areas, such as video retrieval, for which the dataset might be particularly useful. We also provide all the extracted data in order to simplify the use of the dataset.
\end{abstract}



\maketitle

\section{Introduction}
Common datasets form the basis for comparable research by providing a standardized foundation to evaluate methods on and compare their relative behavior. With the ever increasing amount of data produced, especially in the multimedia domain, such datasets also become increasingly larger, especially when they not only cover a highly specialized content area but try to be representative of a larger whole.
The \emph{Vimeo Creative Commons Collection (V3C)}~\cite{rossetto2019v3c} is a video dataset designed to be representative of video content found on the web. The full dataset is split into three shards with a total combined duration of $1\,000$, $1\,300$ and $1\,500$ hours respectively. The first shard (V3C1) has been used since 2019 for several evaluation campaigns, including the Video Browser Showdown (VBS)~\cite{rossetto2020interactive} and the TRECVID~\cite{2020trecvidawad} Ad-hoc Video Search (AVS) Task. While excerpts from the second shard (V3C2) have already been used in recent TRECVID Video-to-Text (VTT) Tasks, the full content of V3C2 is scheduled to enter wider use in 2021. In particular, V3C2 is going to be used for TRECVID AVS 2022 as well as for VBS 2022, and subsequent upcoming iterations of both competitions.  In preparation, analogously to prior work (such as \cite{berns2019v3c1} or \cite{over2009creating}), we present an overview of the properties and contents of this dataset and provide additional resources to facilitate its use. To that end, Section~\ref{sec:overview} provides an overview of the more technical aspects of the dataset while Section~\ref{sec:content} discusses some properties of its content. Section~\ref{sec:conclusion} then concludes the insights and discusses their consequences for the usefulness of the dataset in different research areas. In addition to the presented overview, we make the data generated during this analysis available in order to provide an easier start when using the dataset.\footnote{\url{https://github.com/lucaro/V3C2Analysis}}

\section{V3C2 Overview}
\label{sec:overview}

\begin{figure}[htbp!]
    \centering
    \includegraphics[width=0.46\textwidth]{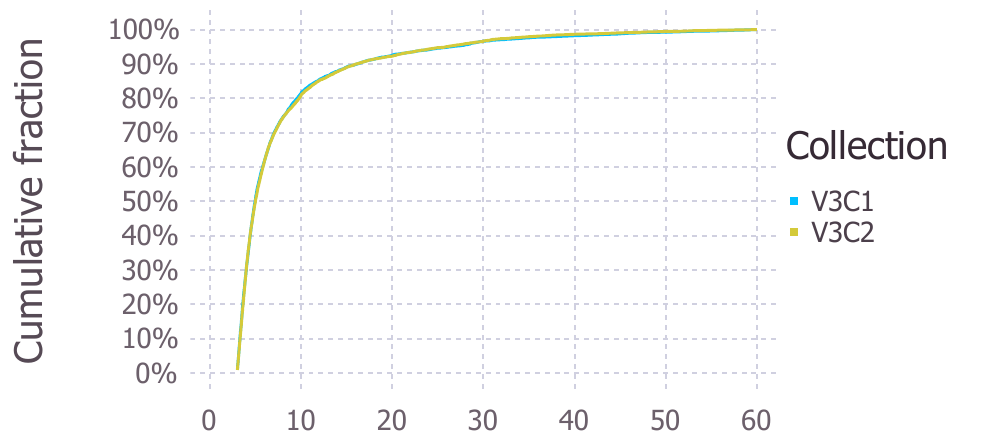}
    \caption{Distribution of video duration in minutes}
    \label{fig:video_duration} %cor of histogram: 0.9387812086998858, cor of cumulative distibution: 0.9998946616361717
\end{figure}

In this section, we analyse several statistics of technical properties of the V3C2 dataset and compare their distribution to those of V3C1. Similarly to the analysis presented in \cite{berns2019v3c1}, we examine the distribution of several properties of V3C2 and compare them with their respective counterparts from V3C1. The dataset is composed of videos collected from the video sharing platform Vimeo\footnote{\url{https://vimeo.com}} which were released under a creative commons license and have a duration between three and 60 minutes. Figure~\ref{fig:video_duration} shows the cumulative distribution of the overall video duration for both V3C1 and V3C2. Both distributions are closely aligned (with a Pearson correlation of 0.939 before cumulative aggregation) and shorter videos clearly dominate both shards, with roughly 80\% of videos having a duration of at most 10 minutes.

\begin{figure}[htbp!]
    \centering
    \includegraphics[width=0.46\textwidth]{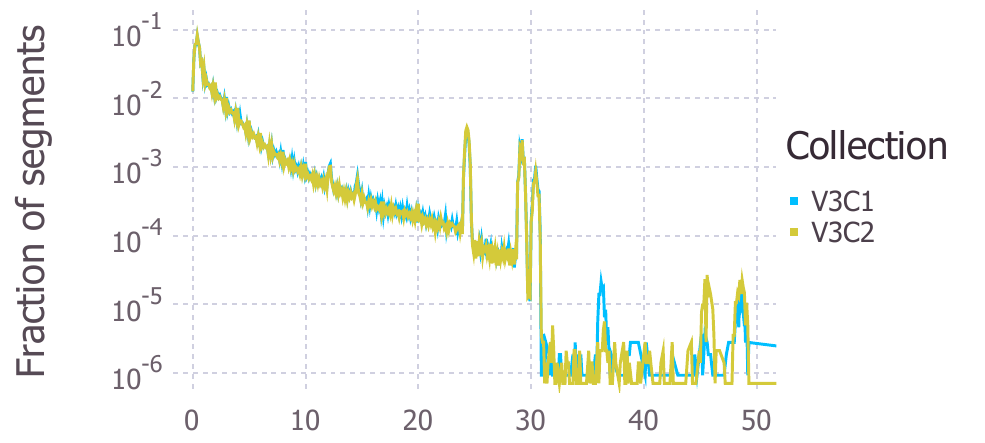}
    \caption{Distribution of segment duration in seconds} %cor: 0.9997891486054277
    \label{fig:segment_duration}
\end{figure}

Figure~\ref{fig:segment_duration} shows the distribution of the duration across the $1\,425\,451$ segments of V3C2, overlaid onto the same distribution from V3C1. It can be seen that both lines have the same overall shape (with a correlation of 0.999) and exhibit only minor differences. Each video in the second shard of the dataset has roughly 146 segments on average, with a minimum of 4 and a maximum of $5\,814$.

\begin{figure}[htbp!]
    \centering
    \includegraphics[width=0.46\textwidth]{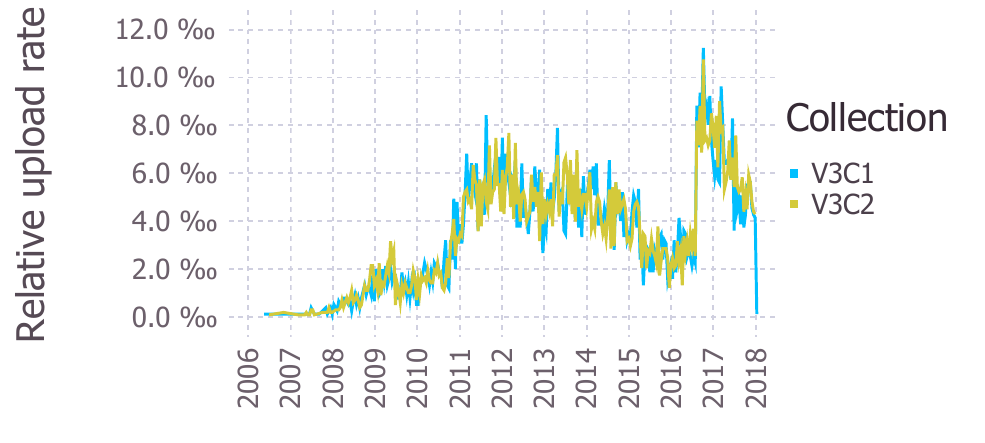}
    \caption{Distribution of video upload over time} %cor: 0.2331602718472932
    \label{fig:upload_date}
\end{figure}

Figure~\ref{fig:upload_date} compares the age distribution of the videos, as indicated by their date at which they have been uploaded to Vimeo, between the two shards of the dataset. The figure shows the upload time-stamps aggregated into two-week intervals. Both curves have a similar overall shape, which is however rather noisy (Pearson correlation coefficient: 0.233).

\begin{figure}[htbp!]
    \centering
    \includegraphics[width=0.46\textwidth]{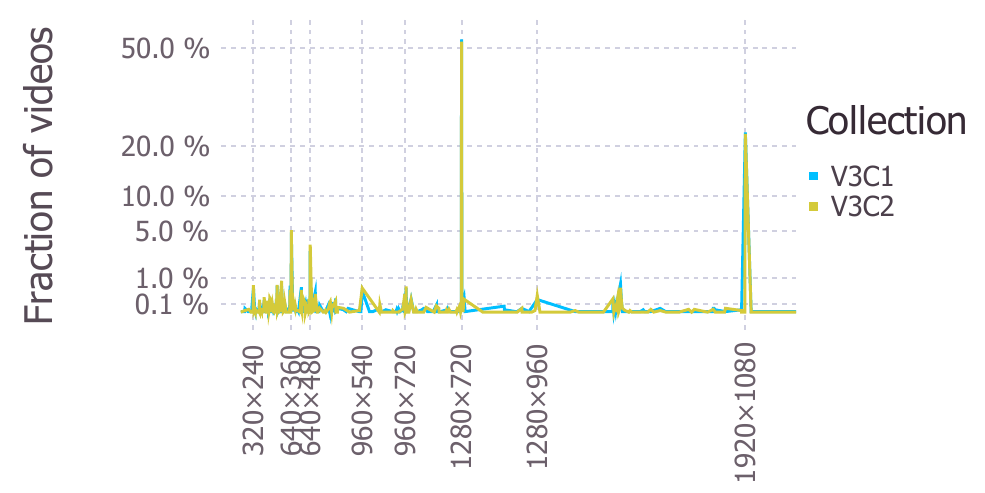}
    \caption{Distribution of video resolution} %cor: 0.48849654342613985
    \label{fig:resolution}
\end{figure}

Figure~\ref{fig:resolution} shows a histogram of the number of pixels per frame, with common resolutions indicated on the horizontal axis. It can be seen that both shards have rather similar distributions (Pearson correlation coefficient: 0.488), with the most commonly occurring video resolution being 1280$\times$720 pixels, followed by 1920$\times$1080.

Since Vimeo preserves at least some of the original uploaded video files and the videos have not been re-encoded when the dataset was compiled, it is not homogeneous with respect to codecs and containers used for encoding. Table~\ref{tab:fileformats} provides an overview of the video containers as identified by their file endings. For both V3C1 and V3C2, the vast majority of videos use MP4 as a container, with the second most popular being MOV, followed by M4V.

\begin{table}
\centering
\caption{Occurrence of file formats}
\label{tab:fileformats}
\begin{tabular}{l|r|r|}
\cline{2-3}
 & \multicolumn{2}{c|}{Count (Fraction)} \\ \hline
\multicolumn{1}{|l|}{Video Container} & \multicolumn{1}{c|}{V3C1} & \multicolumn{1}{c|}{V3C2} \\ \hline
\multicolumn{1}{|l|}{MP4} & 7358 (98.43\%) & 9642 (98.79\%) \\ \hline
\multicolumn{1}{|l|}{MOV} & 102 (1.36\%) & 107 (1.1\%) \\ \hline
\multicolumn{1}{|l|}{M4V} & 11 (0.15\%) & 8 (0.08\%) \\ \hline
\multicolumn{1}{|l|}{AVI} & 1 (0.01\%) & 3 (0.03\%) \\ \hline
\multicolumn{1}{|l|}{MPE} & 3 (0.04\%) & 0 (0\%) \\ \hline
\end{tabular}
\end{table}

\section{V3C2 Content}
\label{sec:content}

\begin{figure}[h]
    \centering
    \includegraphics[width=0.46\textwidth]{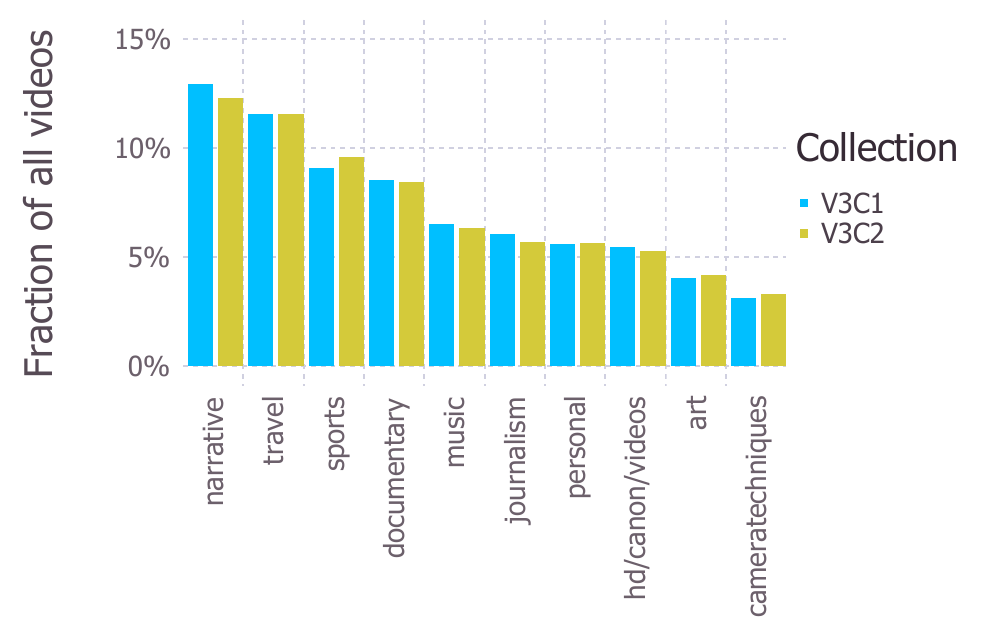}
    \caption{Top-10 Vimeo categories} %corspearman: 0.9871562709590879
    \label{fig:categories}
\end{figure}

A high-level insight into the content of the videos contained within the dataset can be obtained by looking at the categories to which the videos have been assigned by their creators during the upload to Vimeo. Figure~\ref{fig:categories} shows the top-10 assigned category labels for both V3C1 and V3C2. It can be seen that not only are the top-10 categories the same for both datasets, they are also in the same order. Considering all 71 categories that are present in both shards, the Spearman rank correlation between them is 0.987.

\subsection{Low-level Visual}

Table~\ref{tab:colors} shows the dominant colors of the video keyframes, which were made available as part of the dataset. When comparing the distribution of dominant colors across the two shards of the dataset, it can be seen that they are almost identical, despite this not being given any consideration during the creation of the dataset~\cite{rossetto2019v3c}. For both datasets, almost 23\% of keyframes do not have a dominant color, meaning that no color covers at least half the pixels in the frame.
The next largest categories are predominantly gray frames, having a majority of pixels with a saturation of between 2\% and 20\% or a value of 30\% or below in the HSV color space, followed by black and white frames -- which means that the majority of pixels have a color saturation of below 2\%. The most common dominant hues ($h$) with a saturation above 20\% are orange ($0.07 \leq h < 0.14$), red (h < $0.07 \vee h \geq 0.92$) and blue ($0.65 \leq h < 0.73$), followed by green ($0.17 \leq h < 0.44$), cyan ($0.44 \leq h < 0.56$) and magenta ($0.76 \leq h < 0.92$).

\begin{table}[h]
\centering
\caption{Dominant Colors of Shot keyframes}
\label{tab:colors}
\begin{tabular}{l|r|r|}
\cline{2-3}
                                     & \multicolumn{2}{c|}{Key-frame count (ratio)}          \\ \hline
\multicolumn{1}{|l|}{Dominant Color} & \multicolumn{1}{c|}{V3C1} & \multicolumn{1}{c|}{V3C2} \\ \hline
\multicolumn{1}{|l|}{Blue}           & $32\,058$ (2.96\%)            & $40\,024$ (2.81\%)            \\ \hline
\multicolumn{1}{|l|}{Cyan}           & $8\,109$ (0.75\%)             & $11\,032$ (0.77\%)            \\ \hline
\multicolumn{1}{|l|}{Green}          & $10\,912$  (1.01\%)           & $13\,665$ (0.96\%)            \\ \hline
\multicolumn{1}{|l|}{Magenta}        & $1\,899$ (0.18\%)             & $3\,061$ (0.21\%)             \\ \hline
\multicolumn{1}{|l|}{Orange}         & $33\,188$ (3.07\%)            & $48\,683$ (3.42\%)            \\ \hline
\multicolumn{1}{|l|}{Red}            & $30\,569$ (2.82\%)            & $40\,127$ (2.82\%)            \\ \hline
\multicolumn{1}{|l|}{Violet}         & 108 (0.01\%)                  & 152 (0.01\%)              \\ \hline
\multicolumn{1}{|l|}{Yellow}         & $1\,424$ (0.13\%)             & $1\,812$ (0.13\%)             \\ \hline \hline
\multicolumn{1}{|l|}{Black/White}    & $94\,769$ (8.75\%)            & $114\,199$ (8.01\%)           \\ \hline
\multicolumn{1}{|l|}{Gray}           & $624\,835$ (57.71)            & $827\,662$ (58.06\%)          \\ \hline
\multicolumn{1}{|l|}{None}           & $244\,789$ (22.61\%)          & $325\,040$ (22.80\%)          \\ \hline
\end{tabular}
\end{table}

\subsection{High-level Visual}
\label{sec:resnet}
%\todo[inline]{Faces @Klaus?}

\begin{figure}[h]
    \centering
    \includegraphics[width=0.46\textwidth]{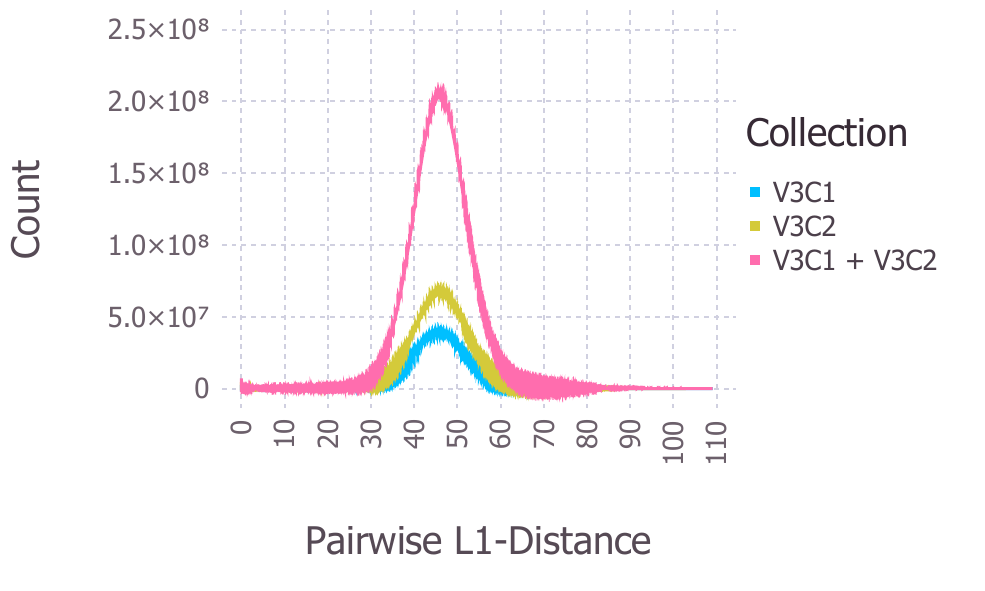}
    \caption{Distribution of pairwise distances of semantic features extracted using an InceptionResNet.}
    %V3C1 mean: 45.917683444097236, std: 6.492919349371129
    %V3C2 mean: 46.22709575581058, std: 6.489823152880049
    %total mean: 46.09621236345112, std: 6.491085271947892
    \label{fig:resnet_hist}
\end{figure}

In order to get a sense of the semantic variability of the dataset, we use the last layer of an InceptionResNet~\cite{szegedy2016inception} which was pre-trained on ImageNet~\cite{deng2009imagenet} to encode the visual semantic information of the videos. The feature representation is generated for each frame and mean-pooled by segment for both V3C1 and V3C2. For the resulting $2\,508\,108$ feature vectors, the pair-wise L1 distance is computed. Figure~\ref{fig:resnet_hist} shows a histogram of the resulting distances for both shards of the dataset individually as well as in combination. 

It can be seen that the three histograms are all rather well aligned (with a $\mu$ of 45.918, 46.227, and 46.096 a $\sigma$ of 6.493, 6.490, and 6.491, respectively), indicating that the semantic variability (as measured by the used neural network) within each collection has a similar distribution than the combination of both shards. 

\subsection{Detected Objects}
\label{sec:yolo}
In order to get an overview of the semantic content in the dataset, we also performed object detction with the YOLOv4 \cite{bochkovskiy2020yolov4} detector (using Darknet-53). For this, we simply used the publicly available model that was pre-trained on 80 classes of MS COCO \cite{lin2014microsoft}.
The process detected a total of $5\,009\,059$ object instances, or roughly 3.5 instances per keyframe on average. This average value is above the median of 2 detected objects per keyframe, since for $334\,221$ or about 23.4\% of all keyframes, none of the 80 object classes were detected. Table~\ref{tab:yolo} shows the top-10 most commonly detected objects and their occurrence.
%and perform statistical analysis of predominant objects in all keyframes, including the total over all keyframes, as well as the average number of detected objects per frame for the corresponding object category.

%total object instances: 5009059
%mean object instances per frame: 3.514014502066713
%median object instances per frame: 2
%top-10 detected objects
%person,3102682
%car,266083
%chair,164530
%book,104849
%tie,95933
%bottle,79078
%cup,69745
%bicycle,61765
%bird,61177

\begin{table}[h]
    \centering
    \caption{Top-10 detected object instances}
    \begin{tabular}{|l|r|c|l|r|}
    \cline{1-2} \cline{4-5}
    Object & Count & \hspace{4mm} & Object & Count \\ \cline{1-2} \cline{4-5}
       Person  & $3\,102\,682$ & &  Bottle & $79\,078$ \\ \cline{1-2} \cline{4-5}
       Car & $266\,083$ & & Cup & $69\,745$ \\ \cline{1-2} \cline{4-5}
       Chair & $164\,530$ & & Bicycle & $61\,765$ \\ \cline{1-2} \cline{4-5}
       Book & $104\,849$ & & Bird & $61\,177$ \\ \cline{1-2} \cline{4-5}
       Tie & $95\,933$ & & Boat & $51\,442$ \\ \cline{1-2} \cline{4-5}
    \end{tabular}
    
    \label{tab:yolo}
\end{table}

\subsection{Faces}

We further applied face detection to every keyframe by using Facenet~\cite{schroff2015facenet}. An overview of the distribution of detected faces is shown in Table~\ref{tab:faces}. While the number of individual faces detected in a keyframe ranges from 0 to 289, the majority of keyframes has no detectable face. For the frames with detections, the majority has only one face. The total number of detected face instances is $1\,352\,749$, or roughly 43.6\% of the number of detected people as shown in the previous section.\footnote{This discrepancy is mostly explained by the fact that the object detector is also capable of detecting people in scenes where their faces are not visible.} 

\begin{table}[h]
    \centering
    \caption{Detected faces per keyframe}
    \begin{tabular}{|l|r|}
    \hline
     Number of Faces    & Keyframe count (ratio) \\ \hline
      0   &         $848\,099$ (59.5\%) \\ \hline
      1   &         $329\,808$ (23.1\%) \\ \hline
      2   &         $114\,864$ (8.1\%) \\ \hline
      3   &         $49\,582$ (3.5\%) \\ \hline
      4   &         $25\,842$ (1.8\%) \\ \hline
      $\geq 5$   &  $57\,257$ (4\%) \\ \hline
    \end{tabular}
    
    \label{tab:faces}
\end{table}

%0 faces: 848099 (59.5%)
%1 face: 329808 (23.1%)
%2 faces: 114864 (8.1%)
%3 faces: 49582 (3.5%)
%4 faces: 25842 (1.8%)
%5+ faces: 57257 (4%)
%most faces detected: 289
%total: 1352749

\subsection{Scene Text}

To analyze the occurrence of text within the videos, we applied a scene text detection and transcription pipeline on all the pre-extracted keyframes of V3C2. The pipeline uses easyOCR\footnote{\url{https://github.com/JaidedAI/EasyOCR}} in order to detect and extract visible text written using the Latin, Persian, and Thai alphabet as well as Simplified Chinese, Devanagari, Japanese, or Hangul characters.

\begin{figure}
    \centering
    \includegraphics[width=0.46\textwidth]{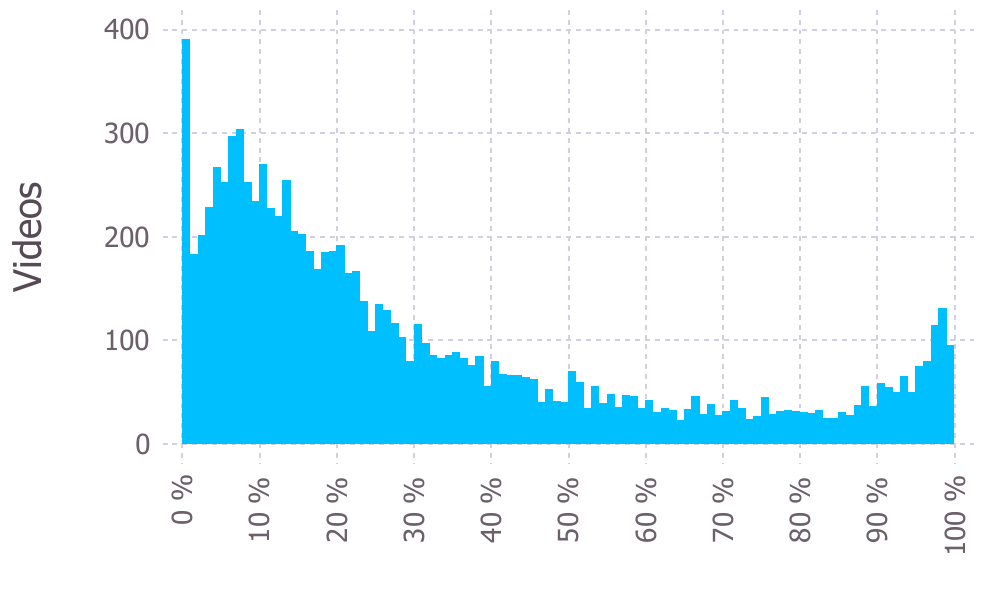}
    \caption{Distribution of fraction of keyframes with text per video}
    \label{fig:text_per_video}
\end{figure}

For the $9\,760$ videos in V3C2, the pipeline identified text in at least one key-frame in $9\,467$ of them. Figure~\ref{fig:text_per_video} shows the distribution of the fraction of keyframes per video. It can be seen that while many videos only have comparatively few segments for which text was detected, the distribution is rather long-tailed, indicating a substantial subset of videos for which text occurs throughout a large fraction of the video. However, of the $1\,425\,451$ segments, only $425\,919$, or roughly 29.88\% contain any detected text. The mean number of keyframes with text per video is roughly 43.64, the median is 21.

\begin{figure}
    \centering
    \includegraphics[width=0.46\textwidth]{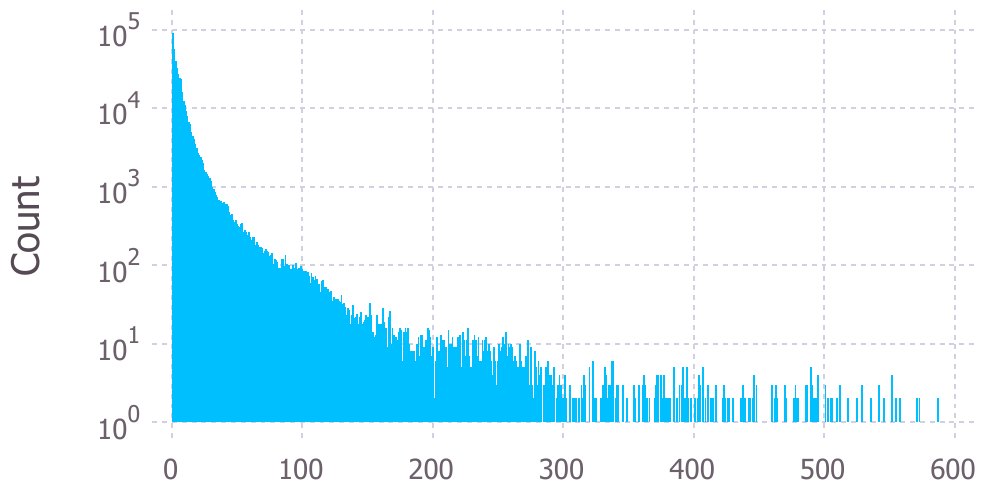}
    \caption{Distribution of detected text items per video}
    \label{fig:text_per_frame}
\end{figure}

Figure~\ref{fig:text_per_frame} shows a distribution of the number of individual detected text items for all frames with any detected text. The used OCR pipeline groups detected text into items if all letters are sufficiently close to each other.
It can be seen that here, the distribution is even more skewed, with the vast majority of the frames only having very few detected items, while only a small number of frames have many text items. For all frames for which any text was detected, roughly 57.43\% have at most 5 detected text items and 77.89\% have 10 or fewer.

\subsection{Speech}

\begin{figure}
    \centering
    \includegraphics[width=0.46\textwidth]{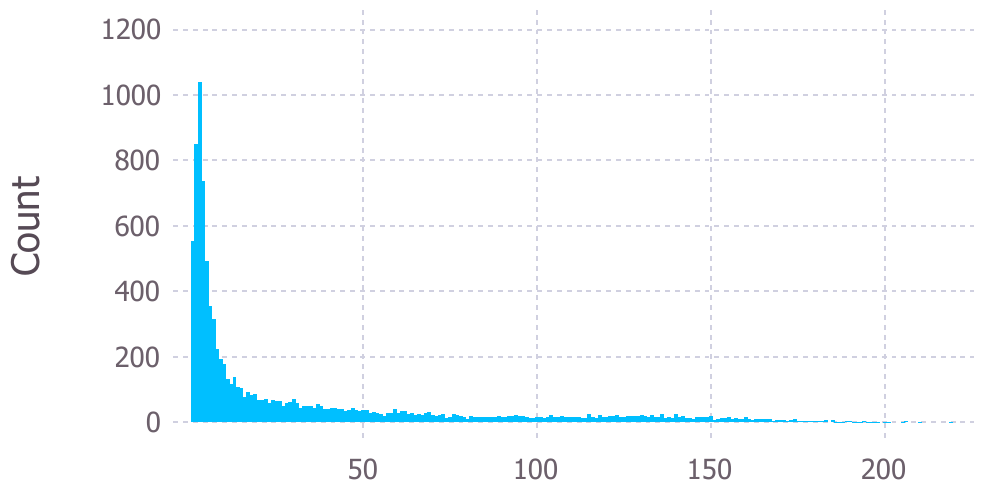}
    \caption{Distribution of detected spoken words per minute}
    \label{fig:words_per_minute}
\end{figure}

In order to analyze the spoken components of the audio found in the dataset, we apply Mozilla's implementation\footnote{\url{https://github.com/mozilla/DeepSpeech}} of DeepSpeech~\cite{hannun2014deep} in version 0.9.3 to the entire second shard of the dataset. The ASR pipeline was prefaced by a voice activity detector\footnote{analogously to the pipeline given in \url{https://github.com/mozilla/DeepSpeech-examples/tree/r0.9/vad_transcriber}} in order to reduce the amount of incorrectly transcribed non-voice audio.
This process produces a list of triples for every video, which contain the start and end timestamp of a segment with detected voice activity as well as the transcribed utterances within this interval. The transcript is always performed under the assumption, that all speech is English.\footnote{This is certainly not true, but in the absence of a reliable method for language identification, it serves as a first-order approximation.} Since these temporal segments are generated only based on voice activity, they are not aligned with the master shot references provided with the dataset. Segments for which voice activity was detected but no transcript could be generated were removed.
The process generated at least one segment with a transcript for $9\,613$ out of the $9\,760$ videos of V3C2. For these videos, voice activity was detected during on average 90.3\% of the video, resulting in 20.225 individual segments per video on average. Figure~\ref{fig:words_per_minute} shows the distribution of the detected number of spoken words per minute for all speech intervals for all videos for which any speech was transcribed.

Both the high fraction of detected speech activity as well as the skewed distribution of words per minute indicate, that the used voice activity detector miss-labeled a sizable fraction of non-speech activity as speech. The figure also indicates that there are only few videos with more or less continuous dialog, while the majority of videos appear to only have occasional spoken words.

\section{Conclusion}
\label{sec:conclusion}
The analyses presented above show that there is a large amount of variation to be found within the dataset --- both on a technical as well as a semantic level and across the modalities. Especially relevant in the area of content-based retrieval, for which the dataset is primarily used, is that none of the methods which generate text can be relied on to index the entire dataset. While text-based queries targeting scene-text or transcripts have been previously shown to be a very effective way of retrieving video sequences (especially in a known-item search setting) neither text generation method covers the dataset to a sufficient degree to be able to rely on them exclusively. 

The large amount of variation also indicates that the dataset could, potentially using some additional annotations, be used for research areas other than retrieval.

With the presented analysis of several aspects of V3C2, we offer insights into its content in order to help researchers estimate if the dataset could be useful for their work. All the generated data on which the presented analysis is based is made available, in order to simplify future use of the dataset.

\appendix

\section{Repository overview}
All the data on which the presented analyses are based is available via \url{https://github.com/lucaro/V3C2Analysis}. The larger resources, including the shot-level features used for the analysis presented in Section~\ref{sec:resnet} as well as the detected object instances discussed in Section~\ref{sec:yolo} are available via \url{https://files.ifi.uzh.ch/ddis/v3c/}.

\begin{figure*}[ht]
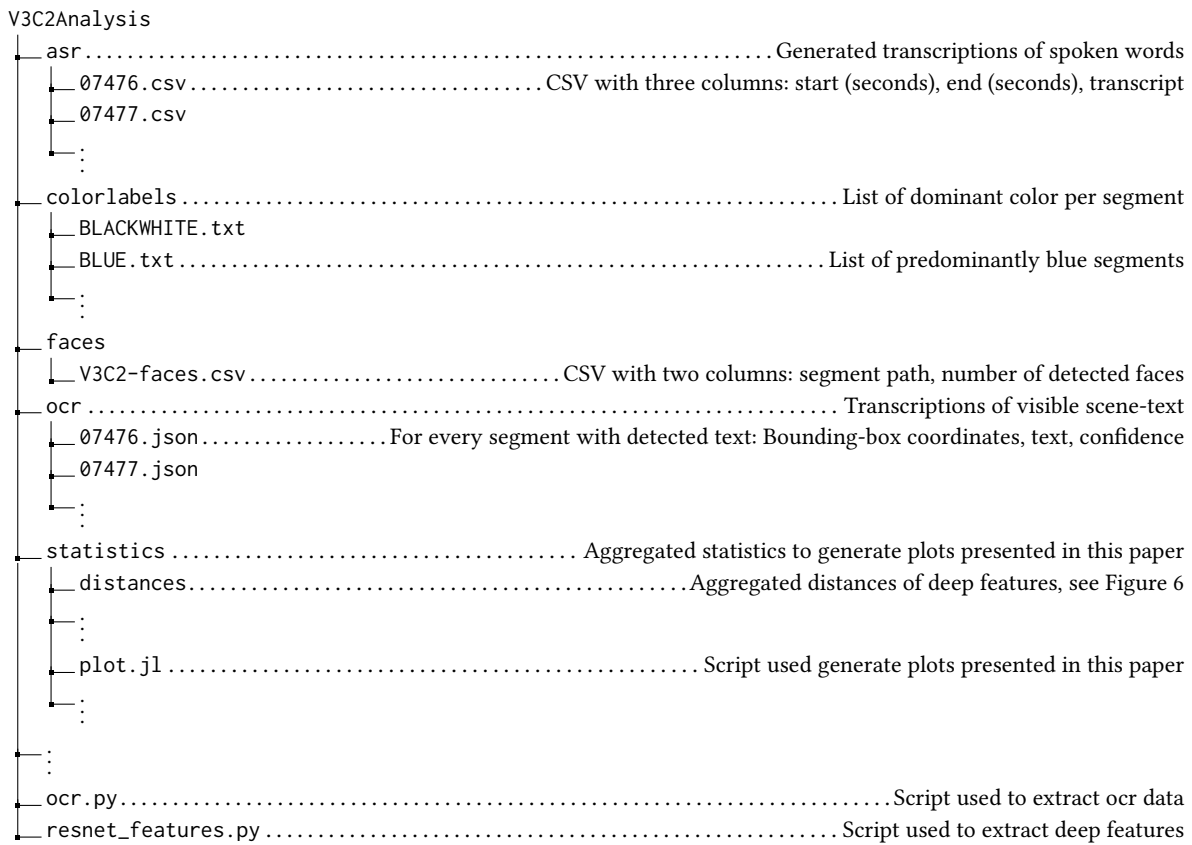

\centering
\begin{minipage}{0.9\textwidth}
  \dirtree{%
  .1 V3C2Analysis.
  .2 asr\DTcomment{Generated transcriptions of spoken words}.
  .3 07476.csv\DTcomment{CSV with three columns: start (seconds), end (seconds), transcript}.
  .3 07477.csv.
  .3 \vdots.
  .2 colorlabels\DTcomment{List of dominant color per segment}.
  .3 BLACKWHITE.txt.
  .3 BLUE.txt\DTcomment{List of predominantly blue segments}.
  .3 \vdots.
  .2 faces.
  .3 V3C2-faces.csv\DTcomment{CSV with two columns: segment path, number of detected faces}.
  .2 ocr\DTcomment{Transcriptions of visible scene-text}.
  .3 07476.json\DTcomment{For every segment with detected text: Bounding-box coordinates, text, confidence}.
  .3 07477.json.
  .3 \vdots.
  .2 statistics\DTcomment{Aggregated statistics to generate plots presented in this paper}.
  .3 distances\DTcomment{Aggregated distances of deep features, see Figure~\ref{fig:resnet_hist}}.
  .3 \vdots.
  .3 plot.jl\DTcomment{Script used generate plots presented in this paper}.
  .3 \vdots.
  .2 \vdots.
  .2 ocr.py\DTcomment{Script used to extract ocr data}.
  .2 resnet\_features.py\DTcomment{Script used to extract deep features}.
  }
\end{minipage}
\caption{Structure of the repository containing the supplementary material}
\label{fig:repo}
\end{figure*}

Figure~\ref{fig:repo} shows an overview of the repository's structure and the data found within. The repository also contains some of the code which was used extract data from the videos and keyframes.

\vspace{1cm}

\bibliographystyle{ACM-Reference-Format}
\bibliography{bibliography}

\end{document}